\documentclass[pre,aps,twocolumn,floatfix,superscriptaddress,10pt]{revtex4-1}

\usepackage{amsmath}
\usepackage{amsfonts}
\usepackage{amssymb}
\usepackage{graphicx}

\begin{document}

\title{Imperfect spreading  on temporal networks}

\author{Martin Gueuning} 
\affiliation{naXys, University of Namur, Rempart de la Vierge 8, 5000 Namur, Belgium}
\author{Jean-Charles Delvenne}
\affiliation{ ICTEAM and CORE, Universit\'{e} catholique de Louvain, Belgium}
 \author{Renaud Lambiotte}
\affiliation{naXys, University of Namur, Rempart de la Vierge 8, 5000 Namur, Belgium}

 \begin{abstract}
 We study spreading on networks where  the contact dynamics between the nodes is governed by a random process and where the inter-contact time distribution may differ from the exponential. We consider a process of imperfect spreading, where  transmission is successful with a determined probability at each contact. We first derive an expression for the inter-success time distribution, determining the speed of the propagation, and then focus on a problem related to epidemic spreading, by estimating the epidemic threshold in a system where nodes remain infectious during a finite, random period of time. Finally, we discuss the implications of our work to design an efficient strategy to enhance spreading on temporal networks.
  \end{abstract}
 
\maketitle

\section{Introduction}

When modelling spreading in systems, it is often assumed that an entity, e.g. a virus or information, performs jumps on a static underlying structure represented by a network  \cite{ep_in_networks}. In its simple form, the network is  assumed to be undirected and unweighted, and it can be made more elaborate by incorporating additional information about the direction of the edges and their weight. In this framework, spreading is described by linear differential equations, where the density on nodes evolves according to the flux induced by neighbouring nodes, and whose solutions is determined by the spectral properties of a matrix encoding the structure of the network, often the Laplacian matrix.
In a large number of systems, however, this modelling approach is not justified because events taking place on the nodes exhibit complex temporal patterns, and the underlying structure has to be considered as a temporal network  \cite{holme2012temporal}. 
For instance, in systems as diverse as mobile phone communication, email checking and brain activity, nodes and links are not permanently active but are punctually activated over time, and their dynamics tends to deviate strongly from that of a Poisson process. 

An important research question associated to the temporality of the network is to understand its impact on the speed and reach of spreading \cite{Liu2013,Masuda2013Temporal,starnini2012random}. For instance, the presence of correlations between  activations of neighbouring edges may either slow-down or accelerate spreading by inducing non-random pathways, depending on the type of correlations  \cite{Scholtes2013SlowDown,vse}. Another important mechanism is associated to the burstiness of the temporal processes, as  two consecutive activations of a link or a node tend to present a broad distribution of inter-contact times  \cite{Karsai12}. The latter mechanism is the subject of this paper. Our main contribution is to consider  a model where spreading is not always successful on an edge when it appears, so that it takes place only after a random number of trials. This process allows to introduce, and to tune,  different time scales in the system, one associated to network evolution, through its inter-contact times, and the other to spreading. As we show analytically and numerically,  imperfect spreading leads to a generalisation of the so-called bus paradox, and has interesting implications on the properties of epidemic spreading on networks and, in particular, on how to optimise the spreading of a spreading process.

\section{spreading on temporal networks}
\subsection{Estimating the inter-success time}

\begin{figure}
		\centering
		\includegraphics[scale=0.65]{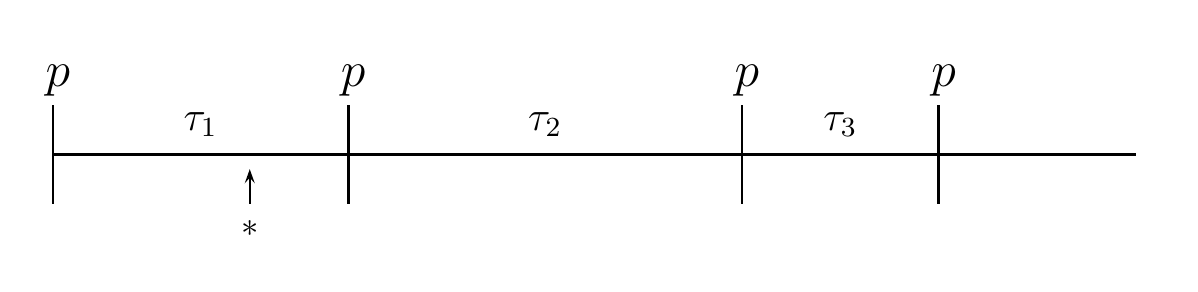}
		\caption{Illustration of the sequence of activation times of an edge. The inter-contact times $\tau_i$ are randomly taken from the distribution $f(\tau)$. The infection of one end of the edge takes place at a random time, illustrated by a star.}
	\label{fig:bus}
\end{figure}

The spreading model is defined as follows. Let us consider a temporal network where the activation of edges is governed by a renewal process, with inter-contact distribution $f(\tau)$. It is further assumed that edges have a vanishing duration, and that the random processes associated to the different edges are independent. spreading takes place from node to node. The diffusing entity, called the virus from now on for the sake of simplicity, sits on a node until an edge appears. The virus then invades the neighbouring node with a probability $p$. In order to describe the random process, it is crucial to estimate the waiting time before a certain edge is invaded, which we call the inter-success time. So far, this model is essentially equivalent to an SI model, which we assume to take place on a tree in order to avoid non-linear effects.
In situations when $p=1$, the relation between the inter-contact distribution and the inter-success distribution $\psi(t)$ is well known to lead to the so-called bus or inspection paradox, as the mean inter-success time $ \left\langle t \right\rangle$ is given by  $ \left\langle \tau^2\right\rangle/2\left\langle \tau\right\rangle$, where $\left\langle \tau\right\rangle$ and $\left\langle \tau^2\right\rangle$ are respectively the first and second moments of the inter-contact time distribution. In situations when the inter-contact distribution presents a broad tail, the average success time can therefore become arbitrarily large as compared to the average contact time. This result, which seems paradoxical at first glance, originates from the fact that the probability for the virus to arrive during a inter-contact time interval is proportional to its duration.
Because the activations of different edges are independent, it can indeed be assumed that a virus arrives at a random time within an inter-contact interval (see Fig. 1). Based on these arguments, one finds that the probability of a success in a first attempt at time $t$ is 
 \begin{eqnarray}
 Q(t) &=&  \frac{1}{\left\langle \tau\right\rangle}\int_t^{+\infty} f(\tau) \, \mathrm d\tau.
  \end{eqnarray}

Let us now turn to the case of an arbitrary value of $p$.
From the previous expression, the inter-success  probability $\psi(t)$ after an arbitrary number of attempts is given by 
\begin{eqnarray}
\psi(t) &=& \displaystyle\sum_{k=0}^{\infty} p (1-p)^{k} P(k+1,t), 
 \end{eqnarray}
 where $p (1-p)^{k}$ is the probability of a success after $k+1$ trials,
\begin{eqnarray}
P(k+1,t)&=& (Q\ast f^{\ast k}) (t),
 \end{eqnarray}
where $\ast$ stands for the convolution product and $P(k,t)$ is the inter-success probability in  $k$ trials. 
This expression takes a simple form in the  associated Laplace space 
\begin{eqnarray}
\tilde{P}(k+1,t)&=& \tilde{Q}(s) \tilde{f}^{k}(s),
 \end{eqnarray}
where the upper tilde stands for the Laplace transform. Using the properties of the small $s$ expansion of each distribution, 
\begin{eqnarray}
\tilde{\psi}(s) &=& 1-s \left\langle t\right\rangle + \frac{s^2}{2}\left\langle t^2\right\rangle + \mathcal{O}(s^3) \cr
\tilde{f}(s) &=& 1-s \left\langle \tau\right\rangle + \frac{s^2}{2}\left\langle \tau^2\right\rangle + \mathcal{O}(s^3),
 \end{eqnarray} 
one directly find a relations between their moments with, in particular for the first moment,
\begin{eqnarray}
	\left\langle t\right\rangle &=& \displaystyle\frac{\left\langle \tau^2\right\rangle}{2\left\langle \tau\right\rangle} + \frac{1-p}{p} \left\langle \tau\right\rangle,
	\end{eqnarray}
	and we recover the expressions of the standard bus paradox when setting $p=1$.

Decreasing $p$ systematically increases the average inter-success time, as expected because more and more trials are required for the spreading to actually take place. In order to account for this trivial effect and to properly identify how the shape of the inter-contact distribution affects $\left\langle t\right\rangle$ as a function of $p$, we focus on the normalised average success time $\bar{\tau}$, a standard measure for the burstiness of a process  \cite{Kivela_Relay,Karsai12}, defined as
\begin{eqnarray}
	\bar{\tau} = \frac{\left\langle t\right\rangle}{\left\langle \tau\right\rangle/p},
\end{eqnarray}
where $\left\langle \tau\right\rangle/p$ is the average success time in the case of a Poisson process. In the latter case, the inter-contact time is exponential and $\left\langle \tau^2\right\rangle = 2\left\langle \tau\right\rangle^2$. One directly finds
\begin{eqnarray}
\bar{\tau}&=& 1+p \left(\displaystyle\frac{\left\langle \tau^2\right\rangle}{2\left\langle \tau\right\rangle^2} - 1 \right),
	\end{eqnarray}
an expression taking the value $1$ in the case of a Poisson processes, as expected, and depending linearly on the so-called burstiness coefficient $$\beta=\frac{\left\langle \tau^2\right\rangle}{2\left\langle \tau\right\rangle^2} - 1,$$ which takes a positive value  in fat-tailed distributions, as the ones observed in a majority of empirical systems. This result clearly shows that burstiness tends to slow down dynamics, but that its impact  becomes less and less important as the probability of success $p$ decreases and a higher number of attempts is necessary for the virus to spread.

\subsection{Epidemic threshold and spreading efficiency}

\begin{figure}
\begin{center}
	\includegraphics[scale=0.55]{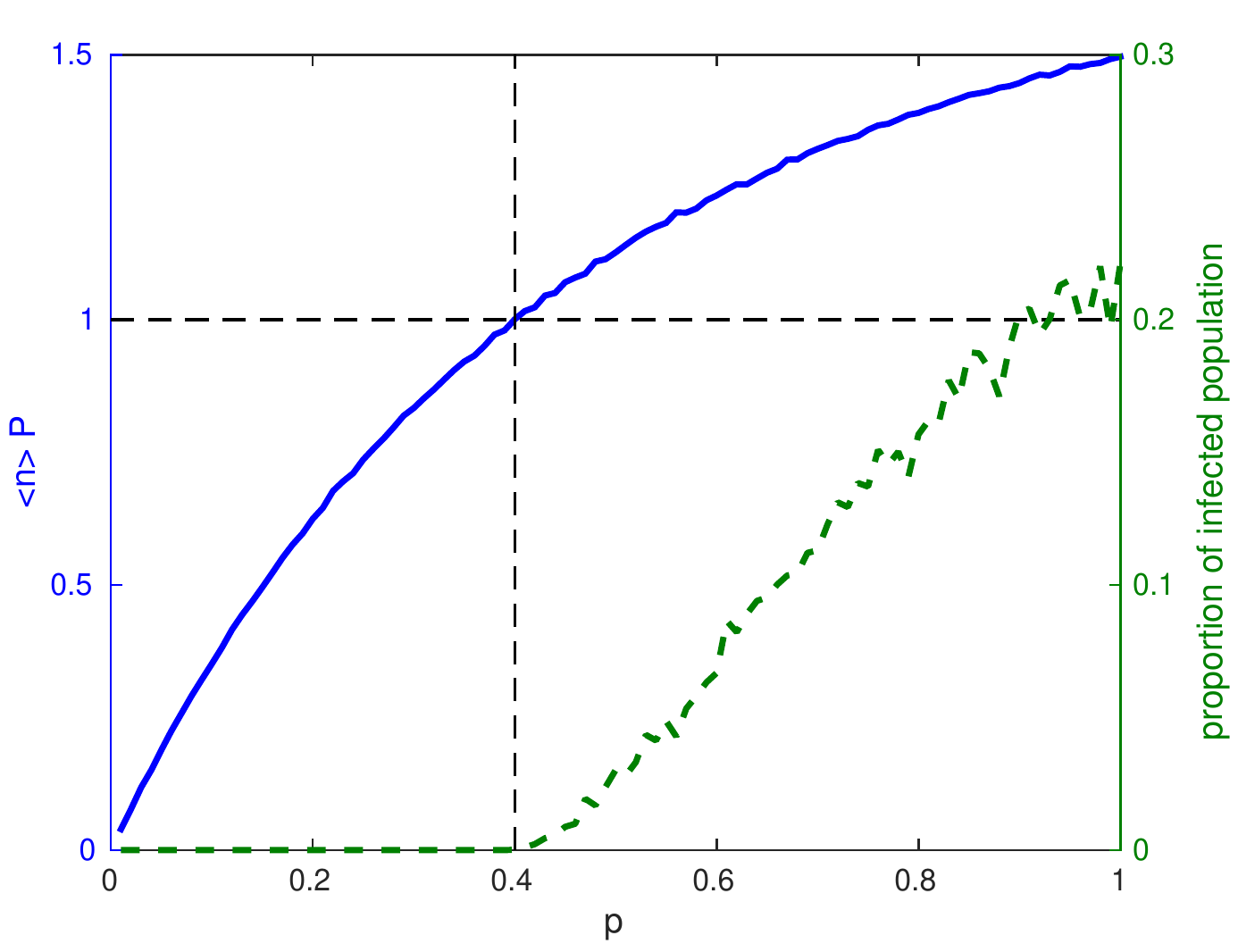} 
\end{center}
\vspace{-0.5cm}
\caption{Proportion of infected population (dashed green) and reproduction number (solid blue) for a tree-like network as a function of $p$. $\left\langle n\right\rangle \mathbb{P}=1$ indicates the epidemic threshold. The parameters of the model are $\left\langle n\right\rangle = 2$, $r(t) = \delta(t-1)$, and the inter-contact distribution is a Gamma distribution, with parameters $a=0.5$ and $b=1$. } 
\label{fig:R0}	
\end{figure}

The dynamics described so far allow to estimate the spread of infection in situations when nodes do not recover from the disease. In a majority of practical situations, however, nodes remain infected only during a finite time before recovering. 
In that case, the propensity of the virus to invade the system is mainly governed by the connectivity of the network, e.g. number of contacts per infected user, and its transmissibility $\mathbb P$, also called infectivity, defined as the probability that the virus spreads to an available neighbour  before the infected node recovers and the virus becomes de-activated. Transmissibility is  defined as
\begin{eqnarray}
\label{trans} 
\mathbb P&=& \int_0^{+\infty} \psi(t) \int_t^{+\infty} r(\tau) \, \mathrm d\tau \, \mathrm dt,
\end{eqnarray}
where, as before, $\psi(t)$ is the probability of a successful infection at time $t$, and $r(\tau)$ is the probability that the infected node recovers at an ulterior time $\tau > t$. This expression clearly shows the importance of the competition between two temporal processes.
In the case of tree-like networks, where all nodes have the same transmissibility $\mathbb{P}$, it is straightforward to show that 
the basic reproduction number, defined as the average number of additional people that a person infects before recovering, is given by \begin{eqnarray}\label{tree}R_0=\mathbb{P} \langle n \rangle,\end{eqnarray} 
where $ \langle n \rangle$  is the 
expected number of susceptible neighbours of an infected node. The epidemic threshold is defined by the condition $R_0=1$ separating between growing and decreasing spreading. The epidemic threshold is thus reduced either by reducing the transmissibility or $ \langle n \rangle$. This result is valid only when the network has a tree-like structure, which is valid for a majority of random network models below the epidemic threshold  \cite{Branching}.

\begin{figure}
\centering
\includegraphics[width=0.43 \textwidth]{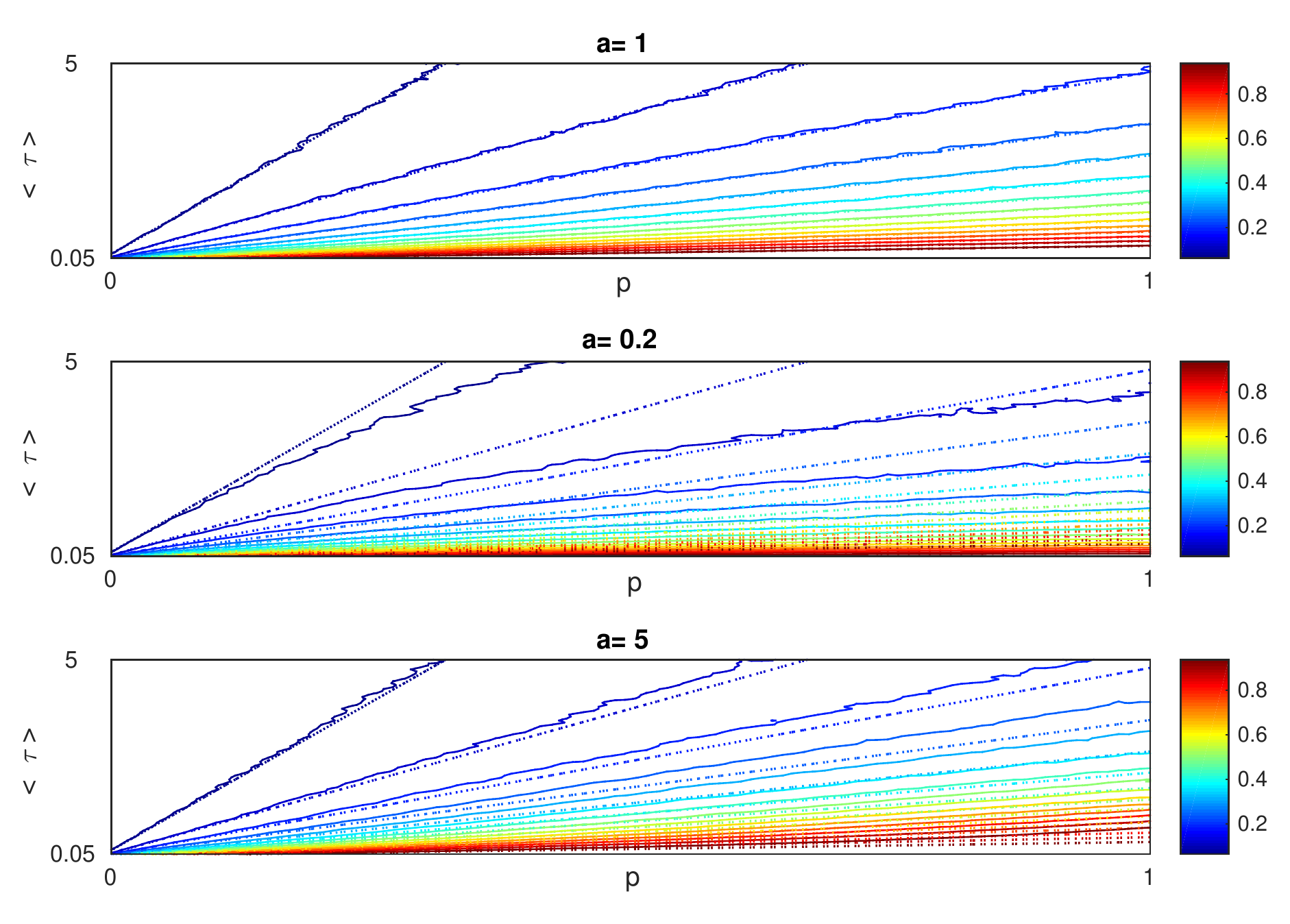}
\caption{Isolines for the transmissibility $\mathbb P$ in the $(\langle \tau \rangle,p)$ plane, for Gamma distributions with different shape parameters $a$ and $r(t) = \delta(t-1)$. In the case of a Poisson process, isolines are linear, indicating a dependency of $\mathbb{P}$ on $\langle \tau \rangle/p$. Dashed lines indicate the linear relationship of a Poisson process. Deviations from $a=1$ alter the convexity of the isolines. }
\label{fig:eff}	
\end{figure}

In order to illustrate these results and to perform numerical tests in the following, we consider an inter-contact distribution given by a Gamma distribution
\begin{center}
\begin{eqnarray*}
f(t;a,b)&=& \displaystyle\frac{1}{b^a \varGamma (a) } t^{a-1} e^{-\frac{x}{b}}
\end{eqnarray*}
\end{center}
with scale parameter $b$ and shape parameter $a$. The family of distributions has the advantage of including the exponential distribution, for $a=1$, and to produce distributions with a tuneable variance by changing $a$. One finds
\begin{center}
\begin{eqnarray}
\label{klkl}
\langle\tau\rangle&=& ab\\
\sigma^2&=&\displaystyle\frac{\langle\tau\rangle^2}{a}
\end{eqnarray}
\end{center}
Fixing $\langle\tau\rangle$, one thus finds that  the variance increases by decreasing $a$. 

Let us also note that transmissibility takes a particularly simple expression when 
 $f$ is an exponential distribution $1/\langle t\rangle e^{-  t/\langle t\rangle}$, so that 
 \begin{eqnarray}
 \mathbb{P} = 1-e^{-p  t_c/\langle t\rangle}
 \end{eqnarray} 
  when the recovery time is a constant $t_c$, $r(t)=\delta(t-t_c)$ and 
   \begin{eqnarray}
  \mathbb{P} = \frac{p t_c / \langle t\rangle}{p t_c/ \langle t\rangle+1}
  \end{eqnarray}
   when the recovery distribution is an exponential with rate $1/t_c$. As a first check, one shows the accuracy of Eq.~\ref{tree} to determine the epidemic threshold on a tree  by numerical simulations in Fig. \ref{fig:R0}.

\begin{figure}

	\includegraphics[scale=0.38]{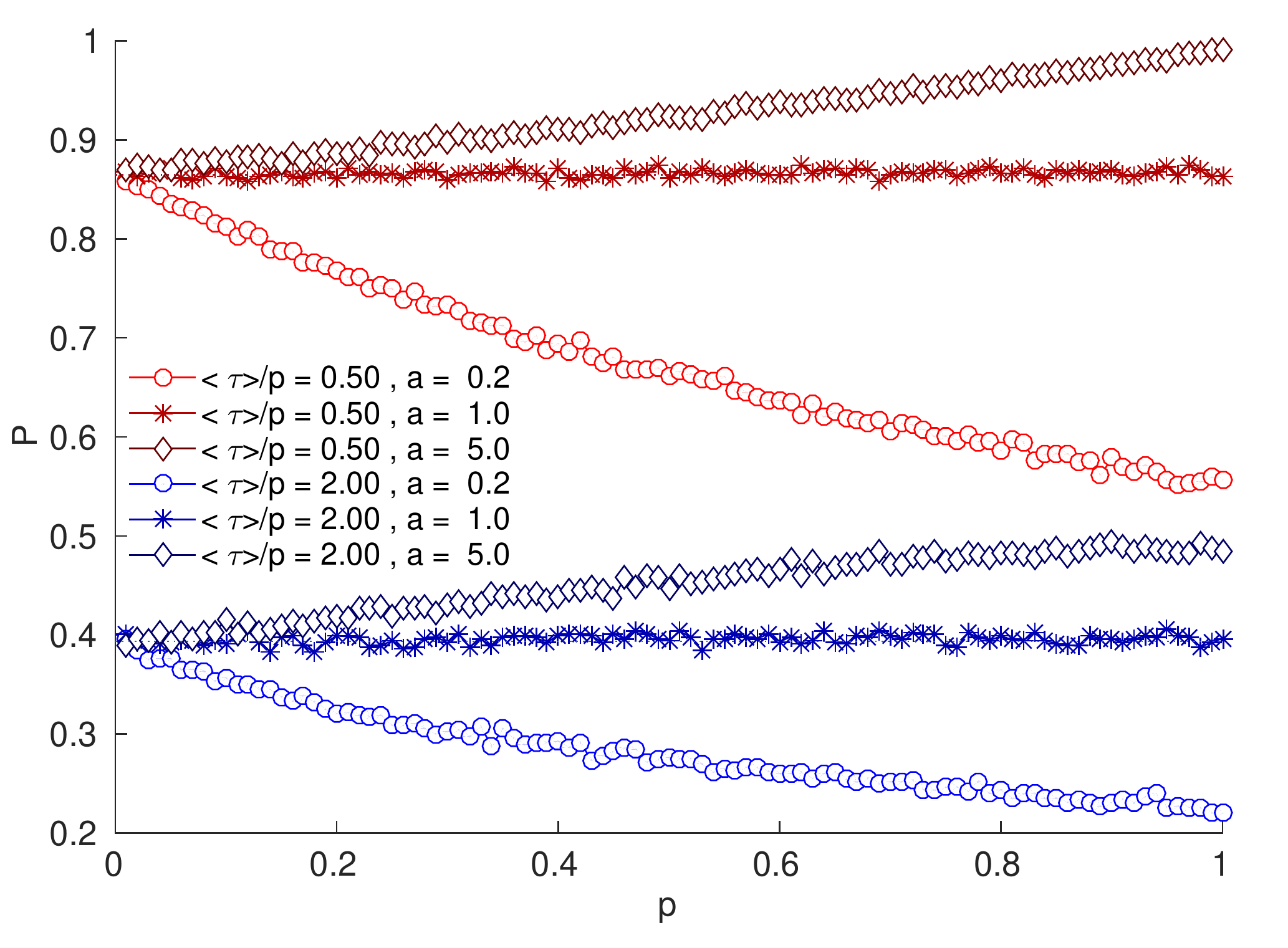} 
\caption{Transmissibility $\mathbb{P}$ vs probability $p$ of success for fixed values for fixed values of $\langle \tau \rangle/p$, with $r(t) = \delta(t-1)$. In the case of Poisson processes, with $a=1$, the system does not exhibit a dependency on $p$. When the system is bursty, in contrast ($a<1$), increasing $p$ decreases the transmissibility of the process. }
\label{fig:ratio_cst}	
\end{figure}

\begin{figure}
 \centering
 	\includegraphics[scale=0.42]{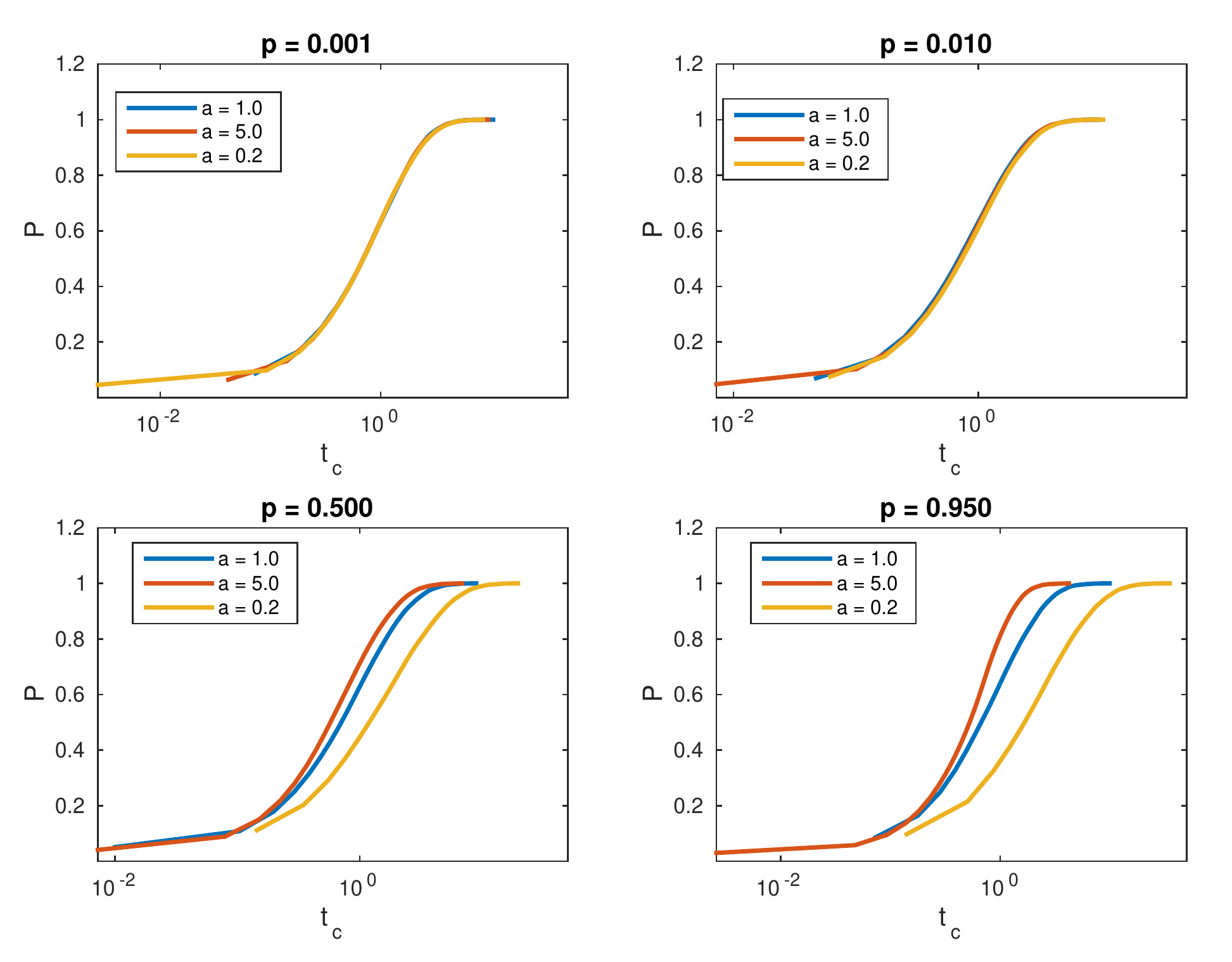} 
 \caption{Transmissibility as a function of the recovery time $t_c$, for $r(t) = \delta(t-t_c)$ and fixed values of $\langle \tau \rangle/p=1$. The dynamics is more and more affected by the shape of the distribution for larger values of $p$. }
 \label{fig:cdf}	
 \end{figure}

Let us now consider the impact of the shape of the Gamma distribution, calibrated by $a$, on transmissibility, by using the exponential case $a=1$ as a baseline. Numerical solutions show in Fig. \ref{fig:eff} the dependency of transmissibility on $\langle \tau \rangle$ and $p$, for three values of $a$ (the value of $b$ is thus determined because of Eq. \ref{klkl}). Simulations are all performed with $r(t) = \delta(t-1)$ but similar results are obtained for exponential distributions for the recovery time. One observes that transmissibility depends on the ratio $\langle \tau \rangle/p$ when $a=1$, but that the system exhibits deviations to this linear relationship when the dynamics deviates from a Poisson process. In order to quantify this deviation, we explore the dependency of $\mathbb P$ on the success probability $p$, for fixed values of the parameter $\langle \tau \rangle/p$. The latter is simply a naive estimation of the time of success, obtained by multiplying the average time between two contacts $\langle \tau \rangle$ and the expected number of contacts before a success takes place $1/p$. Numerical results clearly show, in Fig. \ref{fig:ratio_cst}, that transmissibility becomes less and less efficient, as compared to the Poisson case, for increasing values of $p$, in agreement with the observation that the inter-success time is more and more affected by the bus paradox for increasing values of $p$.
The mechanism behind this effect is also apparent in Fig. \ref{fig:cdf}	where, again for a fixed value of  $\langle \tau \rangle/p=1$, transmissibility is clearly less efficient in the case of bursty dynamics, for larger values of $p$.

This result has important practical implications, for instance in online marketing, where user activity is known to be bursty \cite{web}. Indeed, let us consider a marketing agency having to decide on a strategy in order to optimise its viral impact. The impact of its campaign, evaluated by its transmissibility,  increases only sub-linearly with its quality, evaluated by its success probability $p$. This observation therefore needs to be considered when devising a strategy to balance the gains and the cost of increasing the viral potential of an ad.

\section{Conclusion}

The main purpose of this paper was to study spreading on temporal networks,  where the temporality of edge activations is modelled as a stochastic process and where the inter-contact time distribution takes an arbitrary expression. In practice, we have mainly been interested in situations when the  inter-contact time distribution presents a fat tail, as observed in a variety of social networks. In contrast with a majority of previous works, we have considered a model where the transmission of the diffusive entity is only successful with a certain probability at each contact. Our main results are twofold. First, we have derived an analytical expression for the average inter-success time. We have shown that burstiness plays a more and more important role, associated to a slowing down of the dynamics by the so-called bus paradox, when the probability of success is increased. In the limit of a small probability of success, the shape of the distribution ceases to play a role, and only its average determines the speed of spreading. Second, we have turned to numerical computations in order to calculate the epidemic threshold for a dynamics where nodes remain infectious during a random period of time before recovering. Our results confirm that  the bus paradox hinders the spreading of the process when the probability of success is increased. As we discussed, the results have implications for the design of efficient marketing campaign. In particular, it suggests that an efficient strategy should aim at properly predicting the future activation times of a target user, in order to minimise the time between his infection and his first contact, and therefore the impact of the bus paradox. In this work, we have considered locally tree-like networks, as often assumed when studying spreading processes. An interesting line of future research  would be to incorporate more complex, arbitrary network structures, where cycles and communities play a role \cite{Delvenne2013Bottlenecks}.

\section*{Acknowledgement}
We acknowledge support from IAP DYSCO (funded by the Belgian Science Policy Office) and the ARC `Mining and Optimization of Big Data Models' (funded by the Communaut\'e Wallonie-Bruxelles). Computations were performed at the `plate-forme technologique en calcul intensif' (PTCI) of the University of Namur, Belgium, with the financial support of the F.R.S.- FNRS. 
\section*{Author contribution statement:}
M.G., J.C.D. and R.L. conceived the project, derived the results and wrote the manuscript. M.G. performed the numerical simulations.

\end{document}